\newcommand{\CLASSINPUTtoptextmargin}{1.9cm}
\newcommand{\CLASSINPUTbottomtextmargin}{2.5cm}
\PassOptionsToPackage{bookmarks=false}{hyperref}

\documentclass[10pt,conference]{IEEEtran}

\IEEEoverridecommandlockouts
\usepackage{amsmath,amssymb,amsfonts}
\usepackage{algorithmic}
\usepackage{textcomp}
\usepackage{xcolor}
\usepackage{booktabs}
\usepackage{orcidlink}
\usepackage{pgfplots}
\pgfplotsset{compat=1.18}
\usepackage{graphicx}
\usepackage{caption}
\usepackage{subcaption}
\usepackage{acronym}

\def\BibTeX{{\rm B\kern-.05em{\sc i\kern-.025em b}\kern-.08em
    T\kern-.1667em\lower.7ex\hbox{E}\kern-.125emX}}

\newacro{IID}[IID]{Independent and Identically Distributed}
\newacro{non-IID}[non-IID]{non-Independent and Identically Distributed}
\newacro{IDS}[IDS]{Intrusion Detection System}
\newacro{NIDS}[NIDS]{Network Intrusion Detection System}
\newacro{HIDS}[HIDS]{Host Intrusion Detection System}
\newacro{FL}[FL]{Federated Learning}
\newacro{FedAvg}[FedAvg]{Federated Averaging}
\newacro{AI}[AI]{Artificial Intelligence}
\newacro{ML}[ML]{Machine Learning}
\newacro{DL}[DL]{Deep Learning}
\newacro{PCA}[PCA]{Principal Component Analysis}
\newacro{SVM}[SVM]{Support Vector Machine}
\newacro{NN}[NN]{Neural Network}
\newacro{DNN}[DNN]{Deep Neural Network}
\newacro{MLP}[MLP]{Multi-Layer Perceptron}
\newacro{GRU}[GRU]{Gated Recurrent Unit}
\newacro{CNN}[CNN]{Convolutional Neural Network}
\newacro{RNN}[RNN]{Recurrent Neural Network}
\newacro{RBM}[RBM]{Restricted Boltzmann Machine}
\newacro{DBN}[DBN]{Deep Belief Network}
\newacro{DoS}[DoS]{Denial Of Service}
\newacro{DDoS}[DDoS]{Distributed Denial Of Service}
\newacro{MITM}[MITM]{Man-In-The-Middle}
\newacro{XSS}[XSS]{Cross Site Scripting}
\newacro{IoT}[IoT]{Internet of Things}

\begin{document}

\title{Federated Deep Learning for Intrusion Detection in IoT Networks}

\makeatletter
\newcommand{\linebreakand}{%
  \end{@IEEEauthorhalign}
  \hfill\mbox{}\par
  \mbox{}\hfill\begin{@IEEEauthorhalign}
}
\makeatother

\author{
\IEEEauthorblockN{Othmane Belarbi}
\IEEEauthorblockA{%\textit{Computer Science \& Informatics}\\
    \textit{Cardiff University}\\
    Cardiff, UK \\
    belarbio@cardiff.ac.uk}
\and
\IEEEauthorblockN{Theodoros Spyridopoulos}
\IEEEauthorblockA{%\textit{Computer Science \& Informatics}\\
    \textit{Cardiff University}\\
    Cardiff, UK \\
    spyridopoulost@cardiff.ac.uk}
\and
\IEEEauthorblockN{Eirini Anthi}
\IEEEauthorblockA{%\textit{Computer Science \& Informatics}\\
    \textit{Cardiff University}\\
    Cardiff, UK\\
    anthies@cardiff.ac.uk}
\linebreakand
\IEEEauthorblockN{Ioannis Mavromatis}
\IEEEauthorblockA{\textit{BRIL, Toshiba Europe Ltd.}\\
    Bristol, UK\\
    ioannis.mavromatis@toshiba-bril.com}
\and
\IEEEauthorblockN{Pietro Carnelli}
\IEEEauthorblockA{\textit{BRIL, Toshiba Europe Ltd.}\\
    Bristol, UK\\
    pietro.carnelli@toshiba-bril.com}
\and
\IEEEauthorblockN{Aftab Khan}
\IEEEauthorblockA{\textit{BRIL, Toshiba Europe Ltd.}\\
    Bristol, UK\\
    aftab.khan@toshiba-bril.com}
}

\maketitle

%
% ---- Abstract ----
%
\begin{abstract}
    The vast increase of \ac{IoT} technologies and the ever-evolving attack vectors have increased cyber-security risks dramatically. %Attacks can compromise \ac{IoT} devices to gain access to sensitive data or control them to deploy further malicious activities. 
% The detection of novel attacks often relies upon \ac{AI} solutions.
A common approach to implementing AI-based \acp{IDS} in distributed \ac{IoT} systems is in a centralised manner. However, this approach may violate data privacy and prohibit \ac{IDS} scalability. Therefore, intrusion detection solutions in \ac{IoT} ecosystems need to move towards a decentralised direction. \ac{FL} has attracted significant interest in recent years due to its ability to perform collaborative learning while preserving data confidentiality and locality. Nevertheless, most \ac{FL}-based \ac{IDS} for \ac{IoT} systems are designed under unrealistic data distribution conditions. To that end, we design an experiment representative of the real world and evaluate the performance of an \ac{FL}-based \ac{IDS}. For our experiments, we rely on TON-IoT, a realistic \ac{IoT} network traffic dataset, associating each IP address with a single \ac{FL} client. Additionally, we explore pre-training and investigate various aggregation methods to mitigate the impact of data heterogeneity. Lastly, we benchmark our approach against a centralised solution. The comparison shows that the heterogeneous nature of the data has a considerable negative impact on the model's performance when trained in a distributed manner. However, in the case of a pre-trained initial global \ac{FL} model, we demonstrate a performance improvement of over 20\% (F1-score) compared to a randomly initiated global model.%a pre-trained initial global \ac{FL} model showed a performance improvement of over $20\%$ (F1-score) when compared against a randomly initiated global model.
\end{abstract}

\begin{IEEEkeywords}
Federated Learning, Intrusion Detection System, Internet of Things, Deep Learning, Deep Belief Networks
\end{IEEEkeywords}

%
% ---- Sections ----
%
\section{Introduction}
\label{sec:introduction}

The increasing adoption of \ac{IoT} devices in industrial applications and public infrastructure as well as the ever-evolving attack vectors and threat actors have increased cyber-security risks dramatically. The complexity and heterogeneity of these systems exacerbate the impact of an attack, ranging from data theft to service disruption and human injury, rendering the security of \ac{IoT} devices a significant task~\cite{kumar2023challenges}. 

In an attempt to enhance the security of \ac{IoT} devices, several studies have focused on AI to build more sophisticated threat detection models. Traditional \ac{IDS} solutions base their operations on \ac{ML} models trained centrally in the Cloud and then deployed across multiple devices. Nevertheless, this centralised approach often suffers from increased network overheads and high latency~\cite{Ferrag}, thereby resulting in slow detection of malicious traffic, lack of scalability and unresponsiveness to attacks in the worst case. In addition, centralised data collection can lead to data privacy and secrecy violations. These challenges urge the need to develop effective \acp{IDS} tailored to the requirements of modern \ac{IoT} systems.

In 2017, Researchers at Google introduced \ac{FL}, a distributed learning paradigm that preserves user privacy and data confidentiality~\cite{pmlr-v54-mcmahan17a}. In \ac{FL}, multiple clients collaboratively train a global model under the orchestration of a central server without sharing any raw data. \ac{FL} can be used to implement a distributed \ac{IDS} addressing the aforementioned requirements~\cite{AlMarri2020FederatedML}. However, the authors in~\cite{CAMPOS2022108661} note that most intrusion detection research using \ac{FL} utilises unrealistic data distribution among the participants and inappropriate datasets. 

To address these issues, this paper presents an \ac{FL}-based \ac{IDS} tailored to \ac{IoT} systems, considering the impact of \ac{non-IID} data. In particular, we use the recent TON-IoT~\cite{ToN-IoT} dataset and partition it according to the destination IP address in order to simulate a real-world scenario. This enables the performance evaluation of our \ac{FL}-based \ac{IDS} with realistic \ac{non-IID} data. We implement two \ac{FL}-based \ac{IDS} flavours, one based on \acp{DNN} and one based on our previous work on \acp{DBN} and evaluate their performance. To address the impact of the non-IID dataset on the detection performance, we evaluate the effect of three aggregation methods (FedAvg, FedProx and FedYogi) and also compare the use of pre-trained models against randomly initialised models. % We implement our \ac{FL} experiments relying on the Flower framework.
The experimental results show that starting \ac{FL} from a pre-trained model alleviates the effect of data heterogeneity and improves the stability of global aggregation in \ac{FL}. In addition, FedProx increases the detection performance, followed closely by FedYogi. Nevertheless, the centralised \ac{IDS} performs significantly better than the \ac{FL} approaches. 

The contributions of this work are summarised as follows:
\begin{itemize}
    \item[--] We design realistic \ac{FL}-based \ac{IDS} experiments, relying on real-world network traffic split based on the device IP address. Our set-up results in a non-IID distributed dataset that reflects the heterogeneity of \ac{IoT} networks. 
    \item[--] We implement and evaluate the performance of a \ac{DBN}-based \ac{IDS} and compare it against a \ac{DNN}-based \ac{IDS} and a centralised \ac{IDS} approach.
    \item[--] We evaluate the impact of pre-training in \ac{FL} on the detection performance given our realistic \ac{non-IID} set-up.
    \item[--] We evaluate the effect of three aggregation methods (FedAvg, FedProx and FedYogi) on the performance of \ac{FL}-based \acp{IDS} in non-IID set-ups. 
\end{itemize}

The remaining part of the paper proceeds as follows: In Section \ref{sec:related work}, we highlight recent relevant work on \ac{FL}-based \ac{IDS} in \ac{IoT} systems. Section \ref{sec:methodology} describes our methodology, including data partitioning and pre-processing. In Section \ref{sec:experiments}, we report the results of our experiments. Finally, we conclude the paper and present pathways for future work in Section \ref{sec:conclusion}.

\section{Related Work}
\label{sec:related work}

In this section, we introduce the technical background related to \ac{FL} and analyse the latest research on intrusion detection in \ac{IoT} using \ac{FL}.

\subsection{Federated Learning for IoT systems}
\label{subsec:federated learning}

%Recent research has demonstrated that centralised-based solutions can be problematic in large-scale \ac{IoT} systems. For example, it is simply infeasible to collect data from highly distributed \ac{IoT} devices to the Cloud for training an \ac{IDS}. These operations would increase the network traffic and result in high latency, high communication overhead and data privacy concerns.

\ac{FL} is a distributed learning paradigm where multiple clients collaboratively train a global model under the orchestration of a central server without sharing raw data with a third party~\cite{pmlr-v54-mcmahan17a}. Instead of exchanging their local data with the central server, the clients share only their model updates, hence preserving the data locality and privacy~\cite{pmlr-v54-mcmahan17a}. In principle, the training process consists of two major phases: local training and global aggregation. Initially, the clients download an initial global model from the central server. Upon retrieving it, each client trains an updated model based on their local data. Then, these selected clients send their model updates back to the central server. Finally, the central server aggregates these updates (typically by averaging) and generates a new global model. This process is repeated for multiple iterations until the desired performance is achieved. Since the raw data is not shared for the training in \ac{FL} settings, the leakage of sensitive information is minimised, and data privacy/secrecy is enhanced. Also, the communication overhead is significantly reduced as there is no need for the clients to transmit their raw data to the central server over the network~\cite{pmlr-v54-mcmahan17a}. Hence, \ac{FL} enables the deployment of \ac{IoT} \ac{IDS} more securely and efficiently~\cite{AGRAWAL2022346}.

%In our work, we consider a horizontal cross-device federated learning setup with ten clients. We also explore the impact of pre-training the initial model in a federated fashion, especially in \ac{non-IID} scenario.

\subsection{FL-based IDS for IoT}
\label{subsec:federated learning-based IDS for IoT}

In recent years, the application of \ac{FL} in \ac{IoT} networks has attracted increasing interest among researchers. The characteristics of \ac{FL} seem to offer various benefits for \ac{IoT} security systems. This section presents works that research the use of \ac{FL} for the specific task of intrusion detection in \ac{IoT} networks.

Various intrusion detection datasets have been used to develop \ac{FL}-based \ac{IDS}. The NSL-KDD dataset is used by~\cite{Rahman}, to train a multi-class classification \ac{DNN}. The authors consider three scenarios with different data distributions. The experimental results of the \ac{FL}-based \ac{IDS} showed that the \ac{FL} approach outperforms the distributed learning setting and can reach comparable accuracy with respect to the centralised learning setting. Also, based on the NSL-KDD dataset,~\cite{AlMarri2020FederatedML} uses \ac{MLP} to implement an \ac{FL}-based \ac{IDS}. The proposed approach is based on mimic learning where a teacher model eliminates training data noise/error and soft labels are passed to the student model as regularization to avoid overfitting. In~\cite{ChenZhuo}, the authors propose a new aggregation method named FedAGRU. The latter uses an attention mechanism to prevent the uploading of unimportant model updates that do not benefit the global model. The experiments conducted on the CICIDS2017 dataset indicate that FedAGRU has higher detection accuracy while reducing communication overhead. Finally,~\cite{HEI2020102033} evaluated an \ac{FL}-based Cloud intrusion detection scheme using the KDDCup’99 dataset. The authors explored the utilisation of blockchain technology to secure the aggregation of the model updates. However, such an approach requires resources that are typically not available in the resource-constrained environment of an \ac{IoT} system.

While the majority of the works mentioned above have been designed for \ac{IoT} scenarios, a key limitation is that they are all based on intrusion datasets which do not fully reflect the \ac{IoT}-specific network traffic. Moreover, the majority of the \ac{IDS} datasets were not designed to be used in an \ac{FL} environment. To that end, recent efforts consider \ac{IoT}-specific datasets to develop \ac{FL}-based \ac{IDS}. For example,~\cite{Mothukuri} evaluated an \ac{FL}-based anomaly detection %for \ac{IoT} security 
using the Modbus dataset. The authors used the \acp{GRU} models for identifying the attacks, and the FedAvg aggregation method. The evaluation results show that the suggested approach outperforms the non-\ac{FL} versions in terms of accuracy in attack detection as well as privacy of user data. Furthermore,~\cite{Popoola} presented a federated deep learning method for zero-day botnet attack detection in IoT-edge devices. The adopted method used the N-BaIoT and Bot-IoT datasets to simulate the zero-day botnet attack and the \ac{DNN} model to classify the network traffic. One network traffic class was not included in each partition to reproduce a real-life scenario. The \ac{FL} method was compared with the centralised, distributed and localised methods. The results demonstrated that the \ac{FL}-based \ac{IDS} is most efficient for the detection of zero-day attacks in IoT devices against all other methods. 

However, the aforementioned studies either do not provide data distributions among the clients, or they simulate scenarios where the client's data distribution does not fluctuate, and all data samples follow the same probability distribution. In realistic \ac{FL} scenarios, \ac{non-IID} data is common when data is generated by the end devices. Additionally, recent studies~\cite{LiXiang,Zhao2018FederatedLW} have established that training on \ac{non-IID} data introduces challenges that could deteriorate the performance of the \ac{IDS}% across the entire network
, and lead to convergence issues. On this question, there is a relatively small body of literature that considers the aspect of \ac{non-IID} in their evaluation of \ac{FL}-based \ac{IDS}. 

The researchers in~\cite{LocKedge} proposed LocKedge, a multi-attack detection mechanism for deployment at the edge. The LocKedge system is based on a \ac{DNN} model together with \ac{PCA} for feature engineering.
The suggested method split the dataset into 4 partitions according to the attackers’ IP address. Similarly, the authors in~\cite{Ferrag,REY2022108693} evaluated the impact of \ac{non-IID} data on the development of \ac{FL}-based \ac{IDS} for \ac{IoT} networks. These studies confirmed the finding about the performance reduction of \ac{FL}  in \ac{non-IID} scenarios. However, they do not provide information on the implementation being used and data distribution among the clients. To this end, we designed an experimental set-up with the necessary non-IID characteristics presented in Section~\ref{sec:methodology}.

In most of the aforementioned works, the model's parameters are initialised with random weights. However, a growing body of research~\cite{chen2023on,NguyenMalik} has explored the impact of pre-training and initialisation on \ac{FL}, especially in \ac{non-IID} scenarios, suggesting that starting from a pre-trained model could reduce both the effect of data heterogeneity and the training time required to reach the final model. Our work contributes to the body of knowledge by exploring the impact of pre-training in \ac{FL}-based \ac{IDS}. In addition, we experiment with various aggregation methods to address the impact of non-IID distributed data.

\section{Methodology}
\label{sec:methodology}

In this section, we describe the key processes of our \ac{FL}-based \ac{IDS} for \ac{IoT}. These include the data distribution among multiple \ac{FL} clients, the data preparation, and learning process.

%%%%%%%%%%%% begin %%%%%%% moved from section 4 %%%%%%%%%%%%%%%%%%
\subsection{Dataset}
\label{subsec:dataset}

The selection of an appropriate dataset is a critical component to consider while designing our \ac{FL}-based \ac{IDS} for \ac{IoT} devices. Most \ac{IDS} datasets are unsuitable for \ac{FL} because they cannot be properly disseminated among the clients in a \ac{non-IID} fashion~\cite{REY2022108693}.
% There are multiple \ac{IDS} datasets available; however, the majority of them are unsuitable in the context of \ac{FL}. Their main problem is that they cannot be properly disseminated among the different clients in a non-IID fashion~\cite{REY2022108693}. 
As a result, our suggested approach focuses on \ac{IoT} datasets for \ac{IDS} that can be partitioned based on the destination IP address, namely MedBIoT~\cite{MedBIoT}, BoT-IoT~\cite{Bot-IoT} and TON-IoT~\cite{ToN-IoT}. In particular, we opted for the TON-IoT dataset from the University of New South Wales, Canberra~\cite{ToN-IoT}, as it presents the best ratio between benign and attack traffic compared to other \ac{IoT}-focused datasets. The TON-IoT dataset is based on a realistic testbed that includes heterogeneous data sources and a wide range of attack types, allowing it to give a comprehensive evaluation of \ac{IoT} security systems. Table \ref{tab:class_distribution} describes the TON-IoT dataset data samples.% The dataset has nine types of attacks and is prone to high-class imbalance. 

\begin{table}[t]
\centering
\caption{Statistics of the original TON-IoT dataset.}
\begin{tabular}{p{0.5cm}p{2cm}p{3cm}p{1.5cm}}
\toprule
\bf N° & \bf Label & \bf Total Data Record  & \bf Ratio(\%)\\
\midrule
 1  & Scanning                      & 7,140,161     & 31.96\\
 2  & DDoS                          & 6,165,008     & 27.6\\
 3  & DoS                           & 3,375,328     & 15.11\\
 4  & XSS                           & 2,108,944     & 9.44\\
 5  & Password                      & 1,718,568     & 7.69\\
 6  & Normal                        & 796,380       & 3.56\\
 7  & Backdoor                      & 508,116       & 2.27\\
 8  & Injection                     & 452,659       & 2.03\\
 9  & Ransomware                    & 72,805        & 0.33\\
 10 & MITM                          & 1,052         & 0.0047\\
  % & Total                         & 21,978,631    & 100.0\\    
 \bottomrule
\end{tabular}
\label{tab:class_distribution}
\end{table}

% \subsection{Data Partitioning}
% \label{subsec:data partitioning}
\subsection{On-the-device Data Preparation}
\label{subsec:data preparation}

\begin{table*}[t]
\caption{Description of the class distribution for each client after pre-processing.}
\label{tab:client distribution}
\centering
\resizebox{\textwidth}{!}{\begin{tabular}{llllllllllll}
\toprule
\footnotesize\bf Client & \footnotesize\bf \# Samples & \footnotesize\bf Scan & \footnotesize\bf DDoS & \footnotesize\bf XSS & \footnotesize\bf Pwd & \footnotesize\bf DoS & \footnotesize\bf Normal & \footnotesize\bf Back & \footnotesize\bf Injection & \footnotesize\bf Rans & \footnotesize\bf MITM\\
\midrule
 1  & \bf 3,786,885 & 815       & 3,502,650 & 576     & 26,460  & 192,130 & 16,202  & -       & 48,052 & -      & -   \\
 2  & \bf 2,740,330 & 636,963   & 993,069   & 285,436 & 278,833 & 303,583 & 149,315 & -       & 93,129 & -      & 2   \\
 3  & \bf 2,535,874 & 1,167,320 & 6,906     & 344,136 & 864,611 & 13,108  & 48,046  & -       & 91,740 & -      & 7   \\
 4  & \bf 1,984,149 & 568,501   & 465,525   & 280,915 & 95,029  & 399,568 & 79,285  & -       & 95,323 & -      & 3   \\
 5  & \bf 1,378,579 & 13,382    & 630,127   & 604,691 & 2,883   & 86,721  & 27,304  & 18      & 12,889 & -      & 564 \\
 6  & \bf 1,165,842 & 1,161,124 & 210       & -       & -       & -       & 4,508   & -       & -      & -      & -   \\
 7  & \bf 1,008,113 & 412,493   & 452,559   & -       & 4,444   & 116,463 & 22,154  & -       & -      & -      & -   \\
 8  & \bf 680,480   & 680,446   & -         & -       & -       & -       & 22      & -       & -      & -      & 12  \\
 9  & \bf 603,465   & 336,770   & 3,642     & -       & 37,835  & 181,710 & 21,745  & 133     & -      & 21,629 & 1   \\
 10 & \bf 425,249   & 384       & -         & -       & -       & -       & -       & 423,122 & 3      & 1,737  & -   \\
\bottomrule
\end{tabular}}
\end{table*}

\begin{figure}[t]
    \centering
    \includegraphics[width=0.48\textwidth]{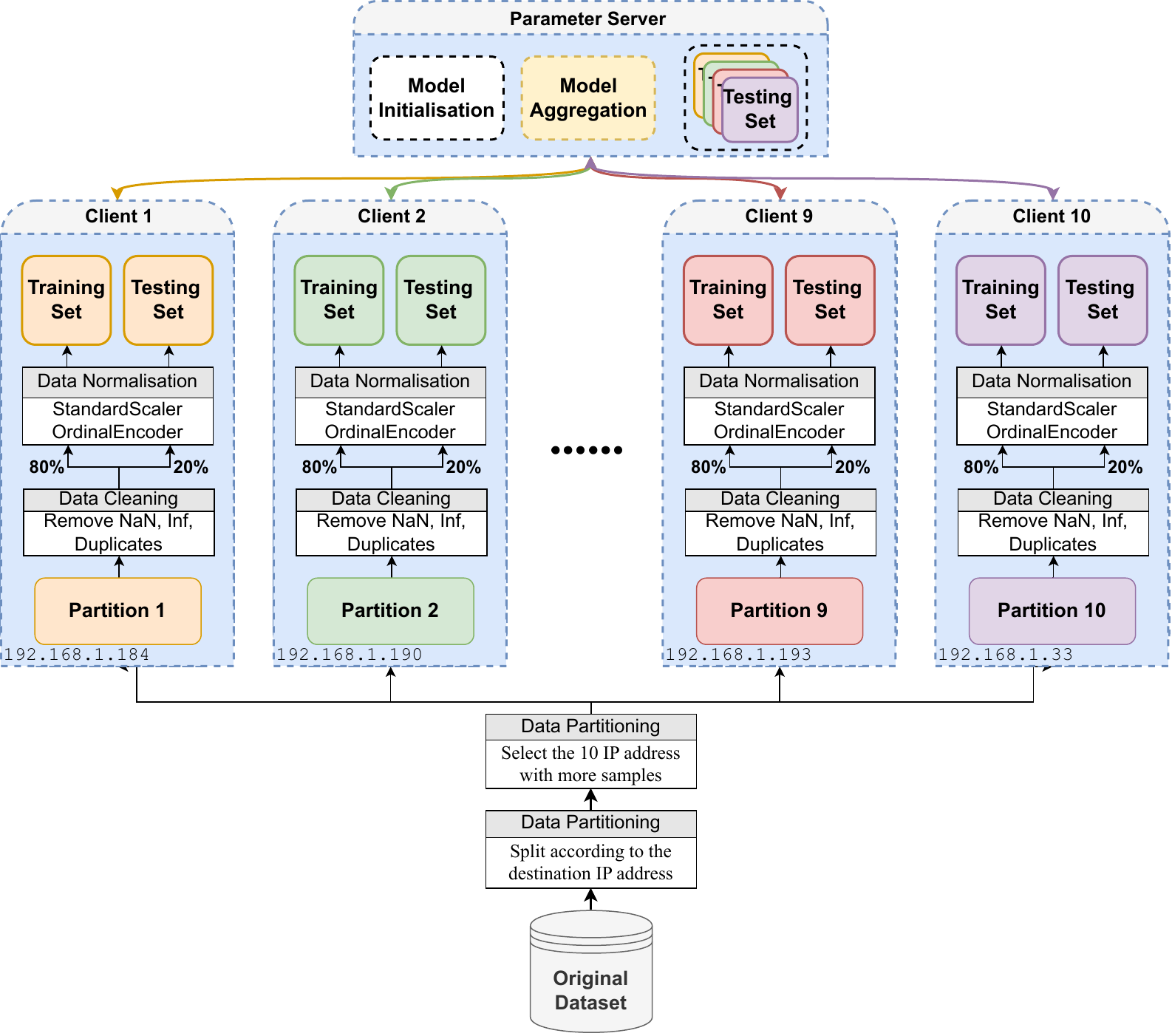}
    \caption{High-level architecture of our \ac{FL}-based \ac{IDS}.}
    \label{fig:flowchart}
\end{figure}

In our data partitioning strategy, we group the data samples by destination IP address, select the top ten IPs with the most samples, and assign each IP's samples to a single FL client. We used only the top ten IP addresses to avoid having clients with minimal data samples. Table \ref{tab:client distribution} describes the data distribution among the ten clients after partitioning and preprocessing. Such distribution allows us to have the most realistic scenario with a high degree of \ac{non-IID}-ness. Moreover, to simulate a realistic scenario where data are generated independently on each client, each \ac{FL} client has to go through data pre-processing individually and independently from the other clients before local model training, as shown in Figure~\ref{fig:flowchart}. 

\subsubsection{Data Cleaning}
\label{subsubsec:data cleaning}

One of the major imperfections in intrusion datasets is missing data in collected samples. To handle missing values, we remove the rows containing NaN values since the dataset is large enough and the fraction of rows containing missing values is small. Hence, it will have minimal effect on performance. Moreover, we eliminate duplicate instances and samples with infinite values for the same reasons. After dropping the unwanted values, we removed unnecessary flow features such as ``\textit{ts}'', ``\textit{src\_ip}'', ``\textit{src\_port}'', ``\textit{dst\_ip}'', and ``\textit{dst\_port}'' in order to ensure that the model learns the complexity of the flow properties instead of the specific endpoints responsible for each traffic, and avoid overfitting. We also encoded the categorical features with ordinal encoding and standardised the numerical ones. We then split the local on-the-device dataset into training and testing sets in the proportions of $80$\% and $20$\%, respectively. A stratified split is crucial in datasets with high-class imbalance since it permits a more accurate evaluation of the model's performance~\cite{Hammerla}. By maintaining the number of samples for each class, a stratified data split effectively ensures enough samples of the minority class in the training set.

\subsubsection{Data Normalisation}
\label{subsubsec:data normalisation}

Most detection algorithms only accept numerical input values. Therefore, we convert the categorical features into numerical features. To increase model performance, we normalise the whole set of numerical features without distorting disparities in the ranges of values. This stage is vital for removing any bias from raw data and optimising model training. It is important to note that in order to preserve privacy, we perform the normalisation on each \ac{FL} client independently from the other clients. Table \ref{tab:client distribution} describes the final dataset on each client after partitioning and preprocessing.

% \subsection{Federated Learning Training}
% \label{subsec:federated learning training}

% The server is in charge of coordinating the federated clients' training efforts. It is responsible for aggregating the model updates transmitted by the clients to generate a new global model. Initialising the weights of the Global model is also the server's responsibility. Once completed, the initial model is distributed to all clients, and the training process can begin. %Each client then trains this model with its own local data, computes a new set of weights, and transmits the model updates to the server. After receiving the updated model parameters, the server combines them using an aggregation function to produce an updated global model and distributes it to the clients. This process is repeated several times until the model is converged.

\subsection{Deep Belief Networks}
\label{subsec:deep belief networks}

% In our experiments, we used a \ac{DNN} and a \ac{DBN} as the basic models for our \ac{FL}. In this section, we briefly discuss the structure of a \ac{DBN}.

% \begin{figure}[h]
%     \centering
%     \includegraphics[width=0.48\textwidth]{images/dbn model.pdf}
%     \caption{DBN architecture.}
%     \label{fig:dbn model}
% \end{figure}

\ac{DBN} is a generative model, composed of stacked layers of \ac{RBM}. An \ac{RBM} is an undirected energy-based model that consists of two layers, namely visible and hidden layer~\cite {Salakhutdinov}. The training process of a \ac{DBN} model requires two phases: unsupervised layer-by-layer training and supervised fine-tuning. During unsupervised pre-training, except for the first and last layer, each \ac{RBM} layer is trained using the contrastive divergence algorithm~\cite{Geoffrey}, and the output of each \ac{RBM} is used as input for the subsequent \ac{RBM} layer. In the second phase, all weights are fine-tuned using back-propagation. \acp{DBN} can establish a meaningful pattern for the attacks which outperform traditional models. For more details on \acp{DBN}, please refer to our previous work in~\cite{Othmane}.

\begin{table}[t]
\caption{Model architecture and hyperparameters.}
\label{tab:model_design_parameters}
\begin{subtable}[t]{0.5\textwidth}
\caption{DBN}
\label{tab:dbn_model_design_parameters}
\centering
\begin{tabular}[t]{p{3.5cm}p{2.1cm}}
\toprule
\bf Attribute                      & \bf Value         \\
\midrule
Input/Output Layer                 & 38:10             \\ 
Hidden Layer                       & 100:150:200:50    \\ 
Hidden Layers Act. Fct.            & ReLu              \\
% Output Layer Act. Fct.             & SoftMax           \\
Optimization Algorithm             & Adam              \\
% Training Loss Function             & Cross Entropy     \\
% Momentum                           & 0.9               \\
Weight init.                       & Xavier initialiser\\
Bias init.                         & Zeros (0)         \\
Gibbs step                         & 1 step            \\
\bottomrule
\end{tabular}
\end{subtable}
\vfill
\begin{subtable}[t]{0.5\textwidth}
\caption{DNN}
\label{tab:dnn_model_design_parameters}
\centering
\begin{tabular}[t]{p{3.5cm}p{2.1cm}}
\toprule
\bf Attribute                      & \bf Value         \\
\midrule
Input/Output Layer                 & 38:10            \\ 
Hidden Layer                       & 128:128:64\\ 
Hidden Layers Act. Fct.            & ReLu              \\
% Output Layer Act. Fct.             & SoftMax           \\
Optimization Algorithm             & SGD              \\
% Training Loss Function             & Cross Entropy     \\
\bottomrule
\end{tabular}
\end{subtable}
\end{table}

\section{Experiments}
\label{sec:experiments}

All the experiments were conducted using an 8-core CPU with 16GB RAM in a macOS environment. The building, training, and evaluation of the \ac{DL} model were performed using PyTorch and the Flower framework.

\begin{figure*}[t]
    \centering
    \includegraphics[width=\textwidth]{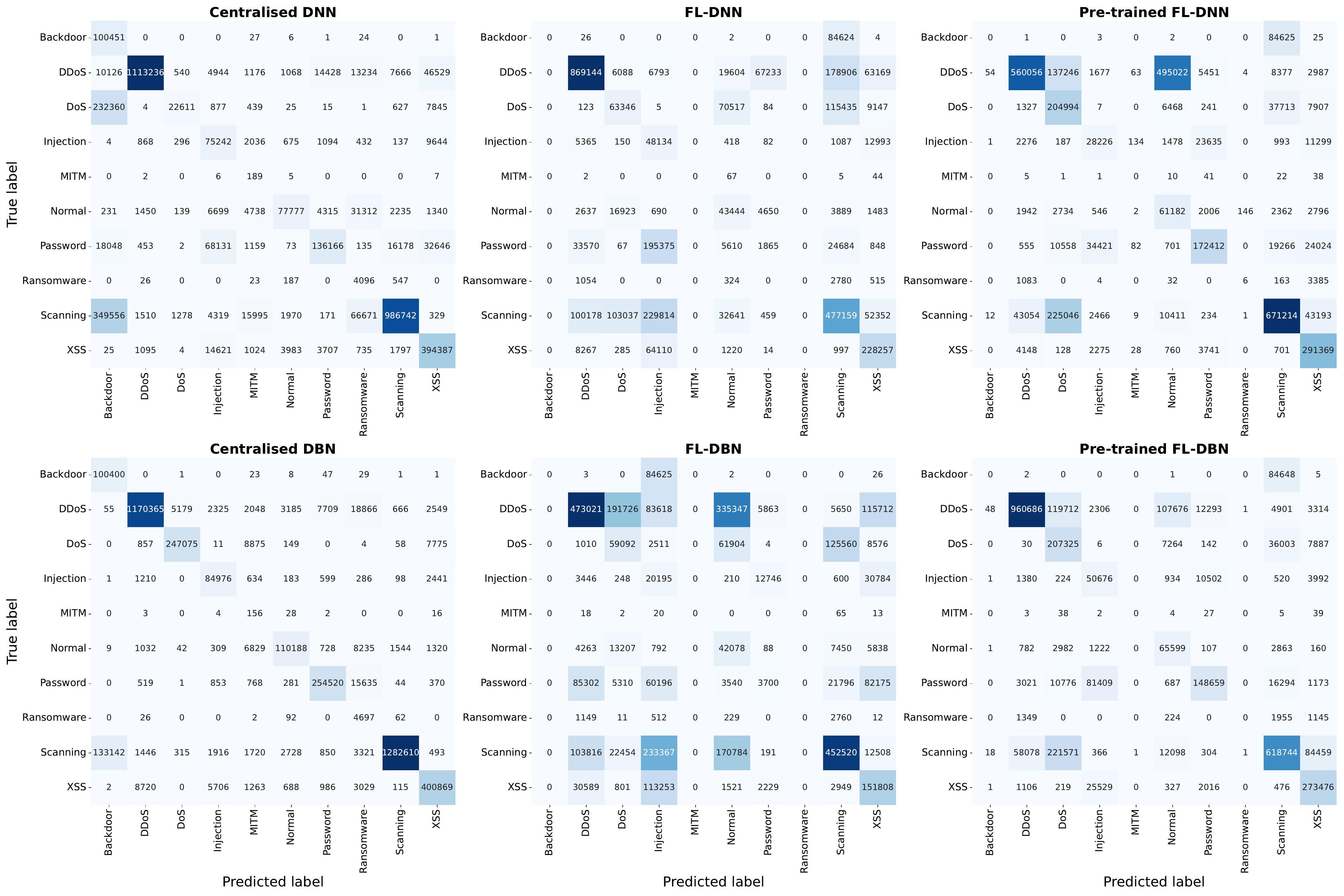}
    \caption{Confusion Matrices.}
    \label{fig:confusion matrices}
\end{figure*}

\subsection{Model Training}
\label{subsec: model training}

% \begin{table}[t]
% \caption{Model architecture and hyperparameters.}
% \label{tab:model_design_parameters}
% \input{tables/dbn architecture and hyperparameters}
% \vfill
% \input{tables/dnn architecture and hyperparameters}
% \end{table}

In this study, we employ a \ac{DBN} and \ac{DNN} for multi-class classification. We train the global \ac{DL} models across 50 rounds and two epochs for each training round, using ten clients with highly skewed data distributions. Since \ac{non-IID} distributed datasets may deteriorate the performance of FedAvg~\cite{LiXiang,Zhao2018FederatedLW}, we experiment with the FedProx~\cite{fedprox} and FedYogi~\cite{reddi2021adaptive} aggregation methods, which are known for their robust convergence in heterogeneous settings~\cite{fedprox,reddi2021adaptive}. Moreover, we initialise the model with pre-trained weights on the server side and compare it against random initialisation. The pre-trained weights are obtained using the data samples from the rest of the destination IP addresses excluded from the experimentation set-up, as described in Section~\ref{subsec:data preparation}. The architecture and hyperparameters of the models are shown in Table \ref{tab:model_design_parameters}.

\subsection{Evaluation Metrics}
\label{subsec:evaluation metrics}

% Depending on the type of dataset and its class distribution, a suitable set of metrics is required to assess a given model's performance reliably. In the case of intrusion datasets that mainly suffer from high-class imbalance, the accuracy metric cannot be relied upon~\cite{Galar}. Therefore, we chose the F1-score, Precision, and Recall as our metrics to evaluate the model. Furthermore, given the unequal distribution of the data based on the label, the weighted average is used in order to consider the size of each class. This is an important consideration, as it ensures that the performance of the model is evaluated fairly, taking into account the distribution of instances in each class.

Depending on the type of dataset and its class distribution, a suitable set of metrics is required to assess a given model's performance reliably. In the case of intrusion datasets that mainly suffer from high-class imbalance, the accuracy metric cannot be relied upon~\cite{Galar}. Therefore, we chose the F1-score, Precision and Recall to evaluate the model. Furthermore, given the unequal distribution of the data based on the label, the weighted average is used to consider the size of each class. It is an important consideration, as it ensures that the distribution of instances in each class is considered.

% \begin{figure}[b]
% \begin{subfigure}[h]{0.24\textwidth}
%     \input{figures/line_chart_dnn.tex}
%     \caption{\ac{DNN}}
%     \label{subfig:comparison dbn}
% \end{subfigure}
% \hfill
% \begin{subfigure}[h]{0.24\textwidth}
%     \input{figures/line_chart_dbn.tex}
%     \caption{\ac{DBN}}
%     \label{subfig:comparison dbn}
% \end{subfigure}
% \caption{F1-score performance for 50 rounds and evaluation of three \ac{FL} strategies (FedAvg, FedYogi, FedProx).}
% \label{fig:aggregation results}
% \end{figure}

% \begin{figure}[t]
% \begin{subfigure}[h]{0.24\textwidth}
%     \input{figures/histogram_dnn.tex}
%     \caption{\ac{DNN}}
%     \label{subfig:comparison dnn}
% \end{subfigure}
% \hfill
% \begin{subfigure}[h]{0.24\textwidth}
%     \input{figures/histogram_dbn.tex}
%     \caption{\ac{DBN}}
%     \label{subfig:comparison dbn}
% \end{subfigure}
% \caption{Precision, Recall and F1-score achieved by the proposed multi-attack classification for the different approaches.}
% \label{fig:comparison}
% \end{figure}

% \begin{figure*}[t]
%     \centering
%     \includegraphics[width=\textwidth]{images/confusion_matrices.pdf}
%     \caption{Confusion Matrices.}
%     \label{fig:confusion matrices}
% \end{figure*}

\subsection{Comparative Results}
\label{subsec:comparative results}

This section presents the experimental results based on the evaluation metrics defined in the previous section. Upon reviewing the literature, no study of \ac{FL}-based \ac{IDS} on \ac{non-IID} data and reproducible methodology was found. Therefore, we could not compare our results against other works. Figure \ref{fig:confusion matrices} shows the confusion matrices of the \ac{DNN} and \ac{DBN} with three different scenarios, namely centralised learning, \ac{FL} and pre-trained \ac{FL}. The correct classifications are represented on the main diagonal of the confusion matrix whereas any off-diagonals denote the erroneously categorised traffic. As a result, the best model will consist of a confusion matrix with only diagonal elements and the rest of the elements near zero. 

The results of the simulations are summarised in Figure \ref{fig:aggregation results}. What stands out in these figures is the clear improvement in performance for the FedYogi and FedProx algorithms compared to FedAvg. These differences can be explained in part by their ability to tackle data heterogeneity. Furthermore, the FedProx simulation presents better convergence stability and prevents divergence through its regularisation hyperparameter.

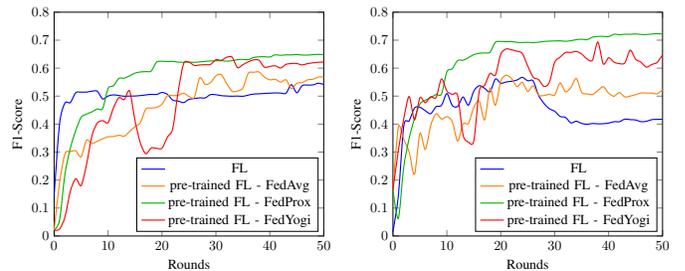
\begin{figure}[t]
\begin{subfigure}[h]{0.24\textwidth}
    \resizebox{\textwidth}{!}{\begin{tikzpicture}
\begin{axis}[
    xlabel=Rounds,
    ylabel=F1-Score,
    xmin=0, xmax=50,
    ymin=0, ymax=0.8,
    xtick={0,10,...,50},
    ytick={0,0.1,...,0.8},
    legend style={at={(0.98, 0.35)},
    legend columns=1},
    every axis plot/.append style={thick}
]

\addplot[smooth,color=blue] coordinates {
(0, 0.14543990864539522)
(1, 0.40098351213854)
(2, 0.4749874113493018)
(3, 0.4767781495199802)
(4, 0.511995296122325)
(5, 0.5143204553968358)
(6, 0.5143356611175877)
(7, 0.5175963311833689)
(8, 0.5174704445754629)
(9, 0.4899957154019418)
(10, 0.500406366028566)
(11, 0.5043933787944519)
(12, 0.5021404706081098)
(13, 0.4996179111420194)
(14, 0.500228762707092)
(15, 0.5004080654452715)
(16, 0.5019694838607675)
(17, 0.5042140797094943)
(18, 0.506434572379696)
(19, 0.5058967387824312)
(20, 0.5130768047458254)
(21, 0.5064379492100306)
(22, 0.4857865849749975)
(23, 0.4823532742368395)
(24, 0.47678691601750156)
(25, 0.4881993893071784)
(26, 0.49127685257181963)
(27, 0.49375575794308535)
(28, 0.4936304372051567)
(29, 0.5054509480079799)
(30, 0.5050262891502103)
(31, 0.5016562513037838)
(32, 0.5004457821769539)
(33, 0.4996030155277795)
(34, 0.500253484758747)
(35, 0.49933534688060954)
(36, 0.5036040397324943)
(37, 0.5073249377458633)
(38, 0.5082471192874528)
(39, 0.5095732298827965)
(40, 0.5105557301577881)
(41, 0.511395223643922)
(42, 0.5110063281524784)
(43, 0.5101575961580586)
(44, 0.5408361127017952)
(45, 0.5118610175157958)
(46, 0.5377253210183481)
(47, 0.5394922566745045)
(48, 0.5400277979070432)
(49, 0.5461969282205019)
(50, 0.5413834596503094)
};
% \addlegendentry{FL-DNN}

\addplot[smooth,color=orange] coordinates {
(0, 0.020888637056476356)
(1, 0.19051469135324217)
(2, 0.29361791120319036)
(3, 0.30172230504509717)
(4, 0.3043287325857423)
(5, 0.28323112066091327)
(6, 0.3424687018610682)
(7, 0.3303395902906872)
(8, 0.3385276323484003)
(9, 0.3474819007965117)
(10, 0.35477286598018976)
(11, 0.35546138191650173)
(12, 0.3603304793683113)
(13, 0.357065153222273)
(14, 0.37350228597744006)
(15, 0.39109718739257215)
(16, 0.4053360591214356)
(17, 0.45372872764614885)
(18, 0.45719866891060945)
(19, 0.459907807058617)
(20, 0.47119457272765736)
(21, 0.5009160554898127)
(22, 0.5017652786748946)
(23, 0.5027001560527146)
(24, 0.510537747283453)
(25, 0.4987732500718536)
(26, 0.4958455442267788)
(27, 0.5655438415234171)
(28, 0.5421474120452834)
(29, 0.5483267594933712)
(30, 0.572405295080062)
(31, 0.5768492008559805)
(32, 0.5447574307960665)
(33, 0.5173575931636297)
(34, 0.518561686715471)
(35, 0.5286559650132726)
(36, 0.5814210363239821)
(37, 0.5851842966622083)
(38, 0.5861969282205018)
(39, 0.5681998859706223)
(40, 0.5582771039670211)
(41, 0.5602772694696364)
(42, 0.5502132592235653)
(43, 0.5581520590391469)
(44, 0.5269642794227289)
(45, 0.5458334313988551)
(46, 0.5482892230102459)
(47, 0.558152)
(48, 0.560277)
(49, 0.5682)
(50, 0.5682)
};
% \addlegendentry{FL-DNN - Pre-trained - FedAvg}

\addplot[smooth,color=black!30!green] coordinates{
(0, 0.020888637056476356)
(1, 0.05639313446102479)
(2, 0.220488299153294)
(3, 0.3134233304323529)
(4, 0.3767020164124989)
(5, 0.4239994030545748)
(6, 0.43568402105995974)
(7, 0.4437019442658499)
(8, 0.4542017959211506)
(9, 0.4556251666604869)
(10, 0.527208569483838)
(11, 0.5370555786906245)
(12, 0.5631766598759614)
(13, 0.5641015346475258)
(14, 0.5725543923815131)
(15, 0.5822043053446018)
(16, 0.5863265257268505)
(17, 0.5875106611622621)
(18, 0.5897374664384999)
(19, 0.6216931256685968)
(20, 0.6244297447527619)
(21, 0.6235712013490889)
(22, 0.6241589693113326)
(23, 0.6214276107083532)
(24, 0.6210329638854061)
(25, 0.621634209782134)
(26, 0.620783938828235)
(27, 0.6220977862169892)
(28, 0.6243952542677053)
(29, 0.6254662177218303)
(30, 0.6260368241869452)
(31, 0.6264958379379119)
(32, 0.624917978759622)
(33, 0.6268590577044881)
(34, 0.6301088339217774)
(35, 0.6313955582214344)
(36, 0.6301421474915934)
(37, 0.6381456566060844)
(38, 0.6403355977504351)
(39, 0.6408508908317847)
(40, 0.6398046449355844)
(41, 0.6475725523515031)
(42, 0.6459659627086769)
(43, 0.6474669758427434)
(44, 0.6473456566632235)
(45, 0.6485066334127643)
(46, 0.6468674482668563)
(47, 0.6466114911540555)
(48, 0.6481968831518912)
(49, 0.6493410395170678)
(50, 0.6480393973429026)
};
% \addlegendentry{FL-DNN - Pre-trained - FedProx}

\addplot[smooth,color=red] coordinates{
(0, 0.020258458259314032)
(1, 0.024738233449082898)
(2, 0.06730523132882751)
(3, 0.1595008849728724)
(4, 0.20457647642088725)
(5, 0.18018174677189377)
(6, 0.25501157016811854)
(7, 0.35731477278467866)
(8, 0.40350589795881997)
(9, 0.41231615738046423)
(10, 0.40523558389588865)
(11, 0.45777202578797577)
(12, 0.49451769362234776)
(13, 0.4875278832400648)
(14, 0.5183161374537321)
(15, 0.4123268358526846)
(16, 0.3320137689886376)
(17, 0.2939888852282332)
(18, 0.31456896435146375)
(19, 0.31354373746620406)
(20, 0.3137739674426466)
(21, 0.3532018890739641)
(22, 0.38408218363789115)
(23, 0.5158599306009182)
(24, 0.6154912159843655)
(25, 0.6214483237463665)
(26, 0.6095151307395101)
(27, 0.6113697051595484)
(28, 0.6122597504041384)
(29, 0.6133138152495798)
(30, 0.6301668677679718)
(31, 0.6280573566849794)
(32, 0.6387137112864822)
(33, 0.6393410395170678)
(34, 0.6019706606573538)
(35, 0.6123573173385891)
(36, 0.6223596153051927)
(37, 0.6294506827700848)
(38, 0.6281809373322368)
(39, 0.6033864274813605)
(40, 0.6059668027550851)
(41, 0.6084623603198995)
(42, 0.6004524417308894)
(43, 0.6134799852731769)
(44, 0.6164744727463524)
(45, 0.616441772460615)
(46, 0.6115899796717106)
(47, 0.6162559878060762)
(48, 0.6189314078225387)
(49, 0.6212007257758118)
(50, 0.6215201636930745)
};
% \addlegendentry{FL-DNN - Pre-trained - FedYogi}
\legend{FL, pre-trained FL - FedAvg, pre-trained FL - FedProx, pre-trained FL - FedYogi}
\end{axis}
\end{tikzpicture}}
    \caption{\ac{DNN}}
    \label{subfig:aggregation dnn}
\end{subfigure}
\hfill
\begin{subfigure}[h]{0.24\textwidth}
    \resizebox{\textwidth}{!}{\begin{tikzpicture}
\begin{axis}[
    xlabel=Rounds,
    ylabel=F1-Score,
    xmin=0, xmax=50,
    ymin=0, ymax=0.8,
    xtick={0,10,...,50},
    ytick={0,0.1,...,0.8},
    legend style={at={(0.98, 0.35)},
    legend columns=1},
    nodes near coords align={vertical},
    every axis plot/.append style={thick}
]
\addplot[smooth,color=blue] plot coordinates {(0, 0.015806497399090187)
 (1, 0.1427677052903542)
 (2, 0.3921783628844793)
 (3, 0.411550383224355)
 (4, 0.45752712957697966)
 (5, 0.4602845958057915)
 (6, 0.4478252470548111)
 (7, 0.4370939253989479)
 (8, 0.462019738360613)
 (9, 0.46719467598547976)
 (10, 0.5091173160596151)
 (11, 0.46662435862777474)
 (12, 0.4629925676031258)
 (13, 0.49671298456634244)
 (14, 0.46766283254825275)
 (15, 0.5181001622644241)
 (16, 0.5372501578300891)
 (17, 0.527869092944105)
 (18, 0.5153474771612357)
 (19, 0.5612166037875199)
 (20, 0.5460579603009337)
 (21, 0.5506539307294911)
 (22, 0.5530866995394942)
 (23, 0.5575007859169799)
 (24, 0.5669931101705562)
 (25, 0.5549011923623226)
 (26, 0.5544998875144332)
 (27, 0.5032790055433612)
 (28, 0.47414254207216366)
 (29, 0.4548923908309082)
 (30, 0.4458744214786647)
 (31, 0.42740072069331775)
 (32, 0.42382975930662053)
 (33, 0.40427332538595356)
 (34, 0.41029769432291585)
 (35, 0.40239317934275953)
 (36, 0.3995055661108968)
 (37, 0.40212875086495753)
 (38, 0.4010685317767103)
 (39, 0.40136694940986667)
 (40, 0.40372279769706937)
 (41, 0.40848565074536897)
 (42, 0.4086014432511711)
 (43, 0.417792957086754)
 (44, 0.41588884439862833)
 (45, 0.41479610339722855)
 (46, 0.40862279042045796)
 (47, 0.4100315745126175)
 (48, 0.41596286178044495)
 (49, 0.4169025163125643)
 (50, 0.4174003177787814)};
\addplot[smooth,color=orange] plot coordinates {(0, 0.1653159116872484)
 (1, 0.3882987856570596)
 (2, 0.3451315311014981)
 (3, 0.3006169452976403)
 (4, 0.22067794010265848)
 (5, 0.36712633260952676)
 (6, 0.3094612453794625)
 (7, 0.39515132079047616)
 (8, 0.43707563970613156)
 (9, 0.4180157509013445)
 (10, 0.42655009573598035)
 (11, 0.40909194643085883)
 (12, 0.33940406647254656)
 (13, 0.40069380632624174)
 (14, 0.4025303473549974)
 (15, 0.46010140555046247)
 (16, 0.48183460022636154)
 (17, 0.4148955080014151)
 (18, 0.5287020651486017)
 (19, 0.4675786178098648)
 (20, 0.5539041045974739)
 (21, 0.5745641266377419)
 (22, 0.5630440602822746)
 (23, 0.5343130178252172)
 (24, 0.5494165096788992)
 (25, 0.5165792408148291)
 (26, 0.5368438753383976)
 (27, 0.5034957091702119)
 (28, 0.5020853695816959)
 (29, 0.5063753713732152)
 (30, 0.5040334460910707)
 (31, 0.5472013801101463)
 (32, 0.515252119088809)
 (33, 0.5308701423127331)
 (34, 0.5634282759782808)
 (35, 0.5135125779010272)
 (36, 0.5076239094357893)
 (37, 0.5059779525328216)
 (38, 0.5195195209724018)
 (39, 0.5336787763535116)
 (40, 0.510469876322883)
 (41, 0.49949725527007793)
 (42, 0.5130311857137989)
 (43, 0.4971106509458317)
 (44, 0.5001902896903713)
 (45, 0.505014960319335)
 (46, 0.5113497782511119)
 (47, 0.5106770299735854)
 (48, 0.5106935179382633)
 (49, 0.508585543173758)
 (50, 0.5231942110019002)};
\addplot[smooth,color=black!30!green] plot coordinates {(0, 0.1653159116872484)
 (1, 0.0628022605152614)
 (2, 0.2455469754667087)
 (3, 0.3490441493898083)
 (4, 0.4195145100103045)
 (5, 0.4721872834955171)
 (6, 0.48519986793525355)
 (7, 0.49412903469961733)
 (8, 0.5058222031203855)
 (9, 0.5074073411134399)
 (10, 0.5871262564680751)
 (11, 0.5980923863597943)
 (12, 0.6271821498783589)
 (13, 0.628212137427427)
 (14, 0.6376256693862721)
 (15, 0.6483723028842975)
 (16, 0.6529630169990844)
 (17, 0.6542817303994093)
 (18, 0.6567616139244454)
 (19, 0.6923490600752262)
 (20, 0.6953966981662346)
 (21, 0.694440580599505)
 (22, 0.695095148873465)
 (23, 0.6920533691216503)
 (24, 0.6916138703631023)
 (25, 0.692283448349872)
 (26, 0.6913365434036283)
 (27, 0.6927997106273417)
 (28, 0.6953582878092198)
 (29, 0.6965509671396226)
 (30, 0.6971864234982083)
 (31, 0.6976976045390177)
 (32, 0.6959404203690219)
 (33, 0.6981021045304233)
 (34, 0.7017212204842556)
 (35, 0.7031541820575162)
 (36, 0.7017583201686456)
 (37, 0.7106714346682825)
 (38, 0.7131102644225616)
 (39, 0.7136841209858814)
 (40, 0.7125189683841651)
 (41, 0.7211697048586578)
 (42, 0.7193805249832311)
 (43, 0.7210521294929543)
 (44, 0.7209170222890274)
 (45, 0.722209944999809)
 (46, 0.7203844651156005)
 (47, 0.7200994182666803)
 (48, 0.7218649913673508)
 (49, 0.7231391820432688)
 (50, 0.721689607166808)};
\addplot[smooth,color=red] plot coordinates {
(0, 0.1653159116872484)
 (1, 0.30536447615239554)
 (2, 0.41881633644383615)
 (3, 0.49862760038978404)
 (4, 0.4152603417335803)
 (5, 0.4940861049685282)
 (6, 0.47625020672837803)
 (7, 0.4978964076250073)
 (8, 0.49492601978875045)
 (9, 0.5613204004344508)
 (10, 0.5180331295865707)
 (11, 0.5061705926943968)
 (12, 0.5016404407484896)
 (13, 0.36155056283088766)
 (14, 0.335762789996182)
 (15, 0.34046268194255613)
 (16, 0.5174022300534425)
 (17, 0.5384892486016811)
 (18, 0.5810914137211389)
 (19, 0.5754326365940728)
 (20, 0.6500483933069572)
 (21, 0.6685441091766144)
 (22, 0.6655218687220225)
 (23, 0.660661459927673)
 (24, 0.6522756762379126)
 (25, 0.6076162705337882)
 (26, 0.5675789447701907)
 (27, 0.5592032848652998)
 (28, 0.5346939551718112)
 (29, 0.5489050810825878)
 (30, 0.6210087144142719)
 (31, 0.6394196529623295)
 (32, 0.6560671191003586)
 (33, 0.6562275859114823)
 (34, 0.6595199276829727)
 (35, 0.6504872446959836)
 (36, 0.6323978497207418)
 (37, 0.6286035411098753)
 (38, 0.6931391820432689)
 (39, 0.6150217260956379)
 (40, 0.6370564793371815)
 (41, 0.6214785251152657)
 (42, 0.6160457975744746)
 (43, 0.6185810524265657)
 (44, 0.6641770560093468)
 (45, 0.6321402867873365)
 (46, 0.6212321367108017)
 (47, 0.6036173392722488)
 (48, 0.6130971795886009)
 (49, 0.6204663459045058)
 (50, 0.6482270811083592)};
% \addlegendentry{FL-DBN}
% \addlegendentry{FL-DBN - Pre-trained - FedAvg}
% \addlegendentry{FL-DBN - Pre-trained - FedProx}
% \addlegendentry{FL-DBN - Pre-trained - FedYogi}
\legend{FL, pre-trained FL - FedAvg, pre-trained FL - FedProx, pre-trained FL - FedYogi}

\end{axis}
\end{tikzpicture}}
    \caption{\ac{DBN}}
    \label{subfig:aggregation dbn}
\end{subfigure}
\caption{F1-score performance for 50 rounds and evaluation of three \ac{FL} strategies (FedAvg, FedYogi, FedProx).}
\label{fig:aggregation results}
\end{figure}

Figure \ref{fig:comparison} depicts the averaged Precision, Recall, and F1-score achieved for the different approaches. On the one hand, we can observe that the centralised approach can detect intrusions with high Precision, Recall and F1-score. It achieves the best results with $97$\% Precision, $93$\% Recall and $94$\% F1-score. However, as mentioned earlier, the centralised approach requires the centralisation of a large amount of data, introducing significant delay at the pre-processing stage and making it easier for intruders to leak sensitive information.

On the other hand, the performance of the \ac{FL}-\ac{DBN} model with random initialisation correctly classified the network traffic with only $54$\% Precision, $37$\% Recall and $42$\% F1-score, while the \ac{FL}-\ac{DNN} model achieved $57$\% Precision, $53$\% Recall and $54$\% F1-score. The model performance reduces significantly when the data is \ac{non-IID} by up to $50$\%. As described by~\cite{Zhao2018FederatedLW}, training on \ac{non-IID} data introduces challenges that deteriorate the performance of the \ac{IDS} and lead to convergence issues. The performance reduction can be explained by the weight divergence of each client. To mitigate this issue, we propose to start \ac{FL} from a pre-trained model to reduce the effect of data heterogeneity and increase the stability of global aggregation in \ac{FL}. We can see that the performance is significantly improved over \ac{FL} with random initialisation. Indeed, the pre-trained \ac{FL}-\ac{DBN} and \ac{FL}-\ac{DNN} models have an F1-score of $72\%$ and $62\%$ respectively. Thus, it demonstrates that the pre-trained model enables the training of more accurate \ac{FL}-based \ac{IDS} (up to $30$\%) than is possible when starting with random initialisation.

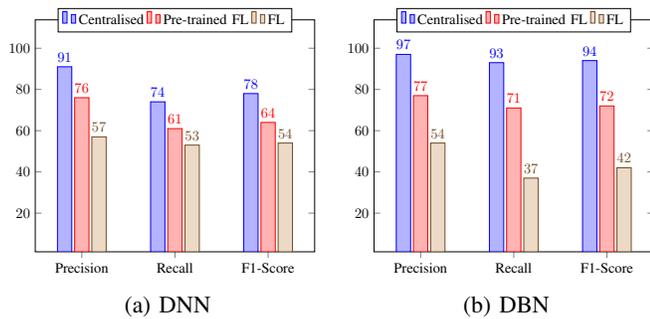
\begin{figure}[t]
\begin{subfigure}[h]{0.24\textwidth}
    \resizebox{0.95\textwidth}{!}{\begin{tikzpicture}
\begin{axis}  
[
    ybar,
    ymin=20,
    ymax=95,
    enlargelimits=0.25,
    symbolic x coords={Precision, Recall, F1-Score},
    xtick=data,
    legend style={at={(0.5, 1.05)},
	anchor=north,legend columns=-1},
    nodes near coords,
    nodes near coords align={vertical},
]
\addplot coordinates {(Precision,91) (Recall,74) (F1-Score,78)};
\addplot coordinates {(Precision,76) (Recall,61) (F1-Score,64)};
\addplot coordinates {(Precision,57) (Recall,53) (F1-Score,54)};
\legend{Centralised, Pre-trained FL, FL}
\end{axis}
\end{tikzpicture}}
    \caption{\ac{DNN}}
    \label{subfig:comparison dnn}
\end{subfigure}
\hfill
\begin{subfigure}[h]{0.24\textwidth}
    \resizebox{0.95\textwidth}{!}{\begin{tikzpicture}
\begin{axis}  
[
    ybar,
    ymin=20,
    ymax=95,
    enlargelimits=0.25,
    symbolic x coords={Precision, Recall, F1-Score},
    xtick=data,
    legend style={at={(0.5, 1.05)},
	anchor=north,legend columns=-1},
    nodes near coords,
    nodes near coords align={vertical},
]
\addplot coordinates {(Precision,97) (Recall,93) (F1-Score,94)};
\addplot coordinates {(Precision,77) (Recall,71) (F1-Score,72)};
\addplot coordinates {(Precision,54) (Recall,37) (F1-Score,42)};
\legend{Centralised, Pre-trained FL, FL}
\end{axis}  
\end{tikzpicture}}
    \caption{\ac{DBN}}
    \label{subfig:comparison dbn}
\end{subfigure}
\caption{Precision, Recall and F1-score achieved by the proposed multi-attack classification for the different approaches.}
\label{fig:comparison}
\end{figure}
\section{Conclusion and Future Work}
\label{sec:conclusion}

The purpose of the current study was to design a multi-class \ac{FL}-based \ac{IDS} for \ac{IoT} and evaluate it under realistic non-IID conditions. Our experimental results confirmed that training an \ac{FL}-based \ac{IDS} on heterogeneous data, either relying on \acp{DNN} or \acp{DBN}, can significantly reduce the model performance by up to 50\%. As a mitigation, we investigated the utilisation of a pre-trained Global model. We demonstrated that pre-training the Global model in \ac{FL} can address data heterogeneity to some extent. In addition, we demonstrated that FedProx and FedYogi increase detection performance since they are designed to address highly non-IID datasets. 

A limitation of this work is that we consider a pre-trained model available. However, it may not always be possible to get public data at the server and pre-train the initial model. As such, exploring pre-training with synthetic data would be a fruitful area for further work. Finally, it is important to note that normalising the data on each client individually can deteriorate the performance of the model. This could be improved by calculating the normalisation parameters centrally, relying on the clients' statistical characteristics. However, sharing the statistical parameters of the clients can disclose sensitive information. Hence, we plan to investigate how we can achieve better normalisation whilst preserving data privacy.

%
% ---- Bibliography ----
%
\bibliographystyle{IEEEtran}
\bibliography{mybibliography}

\end{document}